\documentclass[a4paper,10pt,twoside]{cpc-hepnp}
\usepackage{CJK,upgreek,fancyhdr}
\usepackage{multicol}
\usepackage{graphicx}
\usepackage{booktabs}
\usepackage{amssymb,bm,mathrsfs,bbm,amscd}
\usepackage[tbtags]{amsmath}
\usepackage{lastpage}

\begin{document}
\begin{CJK*}{GB}{gbsn}

\fancyhead[c]{\small Submitted to 'Chinese Physics C'}
\fancyfoot[C]{\small 010201-\thepage}

\footnotetext[0]{Received 21 Junurary 2016}

\title{A method for designing the variable-way high-power cavity combiner\thanks{Supported by National Natural Science Foundation of China (11079034)}}

\author{%
      Yong-Tao LIU (ÁõÓÂÌÎ)
\quad Gui-Rong HUANG (»Æ¹óÈÙ)$^{1)}$\email{grhuang@ustc.edu.cn}\\
\quad Lei SHANG (ÉÐÀ×)
\quad Hong-Xiang LIN (ÁÖºêÏè)
\quad Bai-Ting DU (¶Å°ÙÍ¢)
}
\maketitle

\address{%
National Synchrotron Radiation Laboratory, University of Science and Technology of China, Hefei 230029, China\\
}

\begin{abstract}
The cavity combiner was put forward for high power combining due to its advantages of larger combining ability, variable input ways and less power loss. For the high power cavity combiner, it is better to keep the power loss ratio in a reasonable range, because much power loss would lead to strict requirements on cooling system. The combiner with variable input ways is convenient for outputting different powers according to practical demands. In this paper, a method for designing a variable-way high-power cavity combiner was proposed, based on the relation between input and output coupling coefficients obtained by analyzing the equivalent circuit of the cavity combiner. The method can make the designed cavity combiner in a matching state and its power loss rate in a reasonable range as the number of input ways changes.  As an illustrating example, a cavity combiner with 500 MHz and variable input ways from 16 to 64 is designed, and the simulation results show that our proposed method is feasible.
\end{abstract}



\footnotetext[0]{\hspace*{-3mm}\raisebox{0.3ex}{$\scriptstyle\copyright$}2013
Chinese Physical Society and the Institute of High Energy Physics
of the Chinese Academy of Sciences and the Institute
of Modern Physics of the Chinese Academy of Sciences and IOP Publishing Ltd}%

\begin{multicols}{2}


\section{Introduction}		

The high power Radio Frequency (RF) Solid State Amplifier (SSA) has been widely employed as the power supply in accelerators in the past over ten years \cite{lab1,lab2}. At present the SSA commonly adopts the multi-stage coaxial or waveguide combining structure, whose capability in accommodating input ports on per stage is poor, and the number of inputs is not more than 12 in coaxial or 4 in waveguide normally. In the case with high output power of 100 kW or even more, the SSA's combining tree will be very complicated, and therefore it is inefficient and expensive.

To resolve this problem, the cavity combining technique that can accommodate dozens or even hundreds of input ports, was proposed by the ESRF, and is being developed in recent years \cite{lab3,lab4,lab5}. The condition of an ideal lossless combiner ${N_{loop}}*{\beta _{loop}} = {\beta _{output}} $ is obtained in the design and calculation of the cavity combiner \cite{lab6}. But in our opinion this approximation ignoring the cavity dissipation is, in fact, probably unsuitable. The cavity loss should be considered at first when designing a cavity combiner, since it determines the combining efficiency and the necessity to use the cooling system for outputting high power. In the paper we will analyze the characteristic of the cavity combiner with power loss, derive the relationship between the input and output coupling coefficients and then propose the method for design the cavity combiner to achieve proper combining efficiency and coupling coefficient range. In order to demonstrate the method, numerical study on a cavity combiner as an example of the application of the proposed method is done using the Computer Simulation Technology Microwave Studio (CST MWS) \cite{lab7}.


\section{Equivalent circuit of a cavity combiner}	

The cavity combiner can be considered as a cavity, coupled with $n$ identical input ports and one output port \cite{lab8,lab9}. The simple way to analyze such a RF network is to take the output port as the source end, and the equivalent circuit in this case is shown in Fig. \ref{fig1}, where $I$ is the driving current, ${R_0}$ is the impedance of output port, ${R_S}$ is the cavity shunt impedance, ${R_i}$ is the impedance of input port, and $n_1$, $n_2$  are the transformer ratios of the cavity to input, output coupler, respectively. The coupling coefficients of input and output couplers are defined as
\begin{equation}      
{\beta _i} = \frac{{{R_S}}}{{n_2^2{R_i}}},{\beta _0}  = \frac{{{R_S}}}{{n_1^2{R_0}}}.
\end{equation}
We can simplify the equivalent circuit by normalizing ${R_i}$, and the normalized equivalent circuit is shown in Fig. \ref{fig2}, where the input, output impedances and the shunt impedance of the cavity are represented by 1,  ${{{\beta _i}} \mathord{\left/
 {\vphantom {{{\beta _i}} {{\beta _0}}}} \right.
 \kern-\nulldelimiterspace} {{\beta _0}}}$ and ${\beta _i}$, respectively. Referring to the normalized circuit in Fig. \ref{fig2}, the impedance of the load end is
\begin{equation} 		
{Z_{L0}} = \frac{{\frac{1}{n}{\beta _i}}}{{\frac{1}{n} + {\beta _i}}} = \frac{{{\beta _i}}}{{n{\beta _i} + 1}} ,
\end{equation}
and the reflection coefficient of output port is
\begin{equation} 		
{\Gamma _0} = \frac{{{Z_{L0}} - \frac{{{\beta _i}}}{{{\beta _0}}}}}{{{Z_{L0}} + \frac{{{\beta _i}}}{{{\beta _0}}}}} = \frac{{{\beta _0} - n{\beta _i} - 1}}{{{\beta _0} + n{\beta _i} + 1}}.
\end{equation}
Similarly, specifying an arbitrary input port $i$ as the source end, the load impedance is
\begin{equation} 		
{Z_{Li}} = \frac{1}{{n - 1 + \frac{1}{{{\beta _i}}} + \frac{{{\beta _0}}}{{{\beta _i}}}}} = \frac{{{\beta _i}}}{{{\beta _0} + (n - 1){\beta _i} + 1}},
\end{equation}
and the reflection coefficient of input port is
\begin{equation} 		
{\Gamma _i} = \frac{{{Z_{Li}} - 1}}{{{Z_{Li}} + 1}} =  - \frac{{{\beta _0} + (n - 2){\beta _i} + 1}}{{{\beta _0} + n{\beta _i} + 1}}.
\end{equation}
In general, the output port of combiner must be matched to the transmission line, which means
\begin{equation} 		
{{\Gamma _0} = 0}.
\end{equation}
This is the fundamental characteristic of the perfect multi-way combiner or splitter.

\begin{center}
\includegraphics[width=8cm]{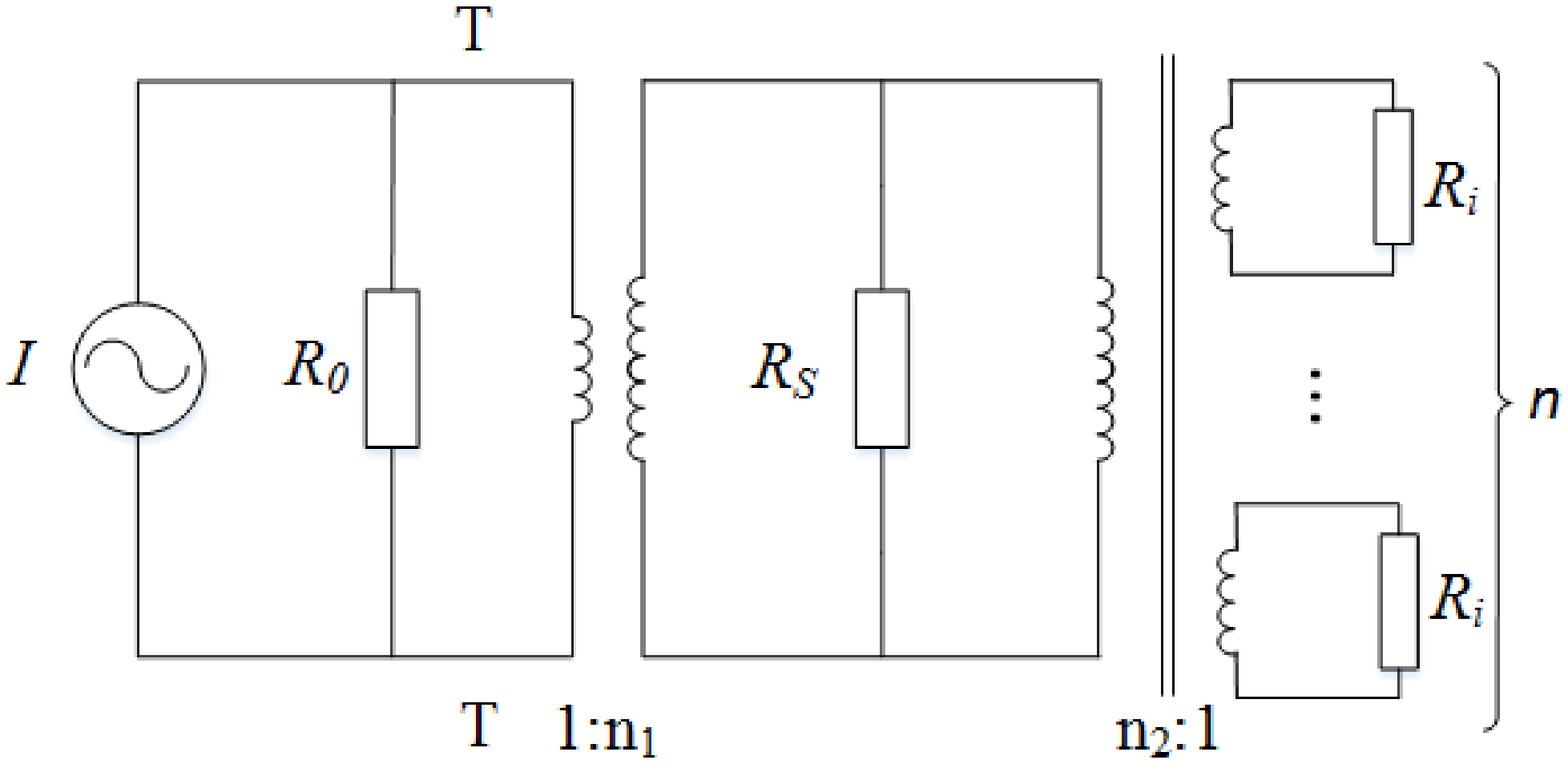}
\figcaption{\label{fig1}   Equivalent circuit of the n-way cavity combiner. }
\end{center}
\begin{center}
\includegraphics[width=8cm]{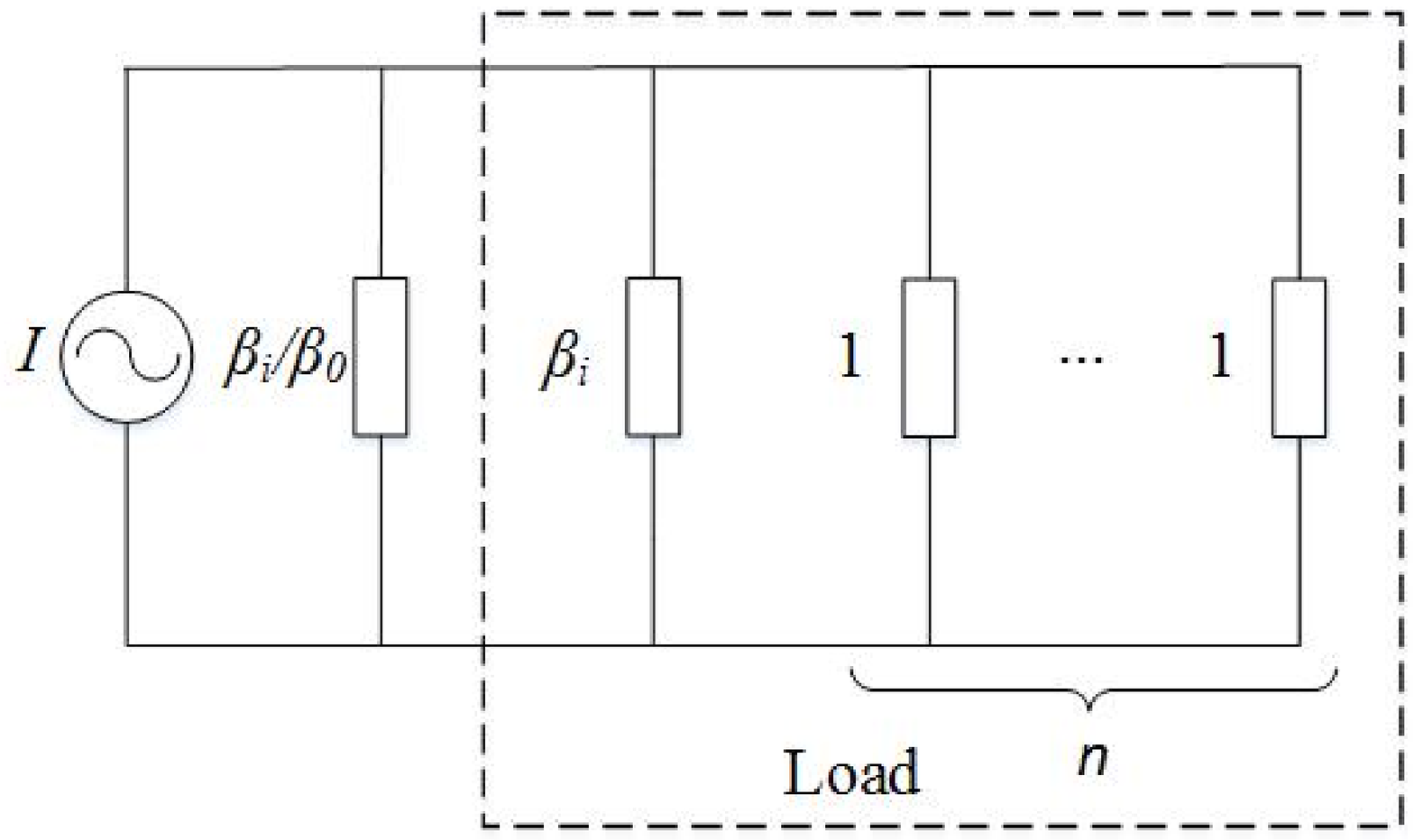}
\figcaption{\label{fig2}   Normalized circuit of the n-way cavity combiner. }
\end{center}

The following derivations are based on the precondition of matching. Therefore, we have
\begin{equation} 		
{\beta _0} = n{\beta _i} + 1,
\end{equation}
\begin{equation} 		
{\Gamma _i} =  - \frac{{(n - 1){\beta _i} + 1}}{{n{\beta _i} + 1}}.
\label{eq:gama}
\end{equation}
According to the equivalent network theory,  the transmission coefficients are derived as
\begin{equation} 		
{S_{i0}} = \sqrt {\frac{{{\beta _i}}}{{n{\beta _i} + 1}}},
\label{eq:Si0}
\end{equation}
\begin{equation} 		
{S_{ij}} = \frac{{{\beta _i}}}{{n{\beta _i} + 1}}{\rm{  }}(i \ne j),
\end{equation}
where $i$ and $j$ ports are different input ports. It can be proved that the relation between the output power ${P _0}$ and the power dissipated in cavity ${P _c}$ is
\begin{equation} 		
{\beta _0} = \frac{{{P_0}}}{{{P_c}}}.
\end{equation}
Then we can define the power loss ratio ${\eta _l}$ and combining efficiency ${\eta _c}$ of the combiner as

\begin{equation} 		
{\eta _l} = \frac{{{P_c}}}{{{P_0}}} = \frac{1}{{{\beta _0}}},
\label{eq:eta}
\end{equation}

\begin{equation} 		
{\eta _c} = 1 - \frac{1}{{{\beta _0}}}.
\end{equation}


\section{Design method}	
A significant advantage of the cavity combiner is that the input couplers can be designed and manufactured in a removable way. So according to the actual output power requirement, we can adjust the amount of input ways, and meanwhile maintain the RF amplifier modules that are connected to the input ways operating in the efficient state. Even though on the occasion that the SSA's output power and the number of the input ways of the cavity combiner are determined, the removable input coupler is still suggested. The reason is that on the design and test stage of the combiner, a prototype is usually needed to be manufactured. It will be convenient for carrying out the RF measurement and the power testing of the prototype, which is coupled partial input couplers.

We redefine the input and output coupling coefficients as ${\beta _i}(n)$ and ${\beta _0}(n)$, respectively. Their relation is
\begin{equation} 		
{\beta _0}(n) = n{\beta _i}(n) + 1,
\end{equation}
where $n$ is the number of inputs. If the maximum amount of input ways is set to $N$, we have
\begin{equation} 		
{\beta _0}(N) = n{\beta _i}(N) + 1.
\end{equation}
When outputting high power, the combining efficiency is important for the cavity combiner. For example, in the case of 100kW output, it is better to set the loss rate ${\eta _l}$ to be less than 1\%. Because in this case the air conditioning can well meet the cavity cooling requirements, or else the water-cooling system probably has to be used.

According to Eq.~(\ref{eq:eta}), ${\eta _l}$ is inversely proportion to ${\beta _0}$. The larger ${\beta _0}$ is, the higher the combining efficiency ${\eta _c}$ becomes. But if the ${\beta _0}$ is larger than 200, it is difficult to be realized due to a strong perturbation of the electromagnetic field caused by the output coupler. Therefore, the ${\beta _0}(N)$  is suggested in the range of 100 to 200, which corresponds to the ${\eta _l}$ in the range of 0.5\%-1\% for a cavity combiner with 100 kW or more output power. In addition, ${\beta _i}(N) < 1$ (under-coupling state) is also unsuitable. In this case the geometrical dimensions of the input coupler are small, so it is hard to maintain the coupling coefficient stable, due to that the coupling coefficient is easy to be changed even the coupler with a small mechanical deviation or deformation. Therefore, when designing a N-way cavity combiner with the maximum output power of 100kW or more, the range of the parameters are suggested as
\begin{equation} 		
\left\{ \begin{array}{l}
100 \le {\beta _0}(N) \le 200\\
{\beta _i}(N) \ge 1
\end{array} \right..
\end{equation}
The upper limit of N is
\begin{equation} 		
N = \frac{{{\beta _0}(N) - 1}}{{{\beta _i}(N)}} \le \frac{{200 - 1}}{1} = 199.
\end{equation}
If the number of the input ways decreases from $N$ to $n$, there are two methods to maintain the cavity combiner under the matching condition.

The first method is to keep ${\beta _0}(n) = {\beta _0}(N)$. This method has the advantage of maintaining a stable combining efficiency of the cavity combiner. But the input coupling coefficient will increase as the amount of inputs $n$ decreases,
\begin{equation} 		
{\beta _i}(n) = \frac{N}{n}{\beta _i}(N).
\end{equation}
Due to that a large amount of inputs are mounted on the cavity, it will take a lot of work to tune the input couplers when the input amount decreases. Besides, it will increase the electromagnetic field perturbation, because the geometrical dimensions of loops are very large if $n$ is small.

The second method is to keep ${\beta _i}(n) = {\beta _i}(N)$, and the output coefficient is changed as
\begin{equation} 		
{\beta _0}(n) = n{\beta _i}(N) + 1.
\end{equation}
The advantage of this method is that we can only tune the output coupler to maintain the cavity combiner matching. With this approach, it will change the combining efficiency of the cavity combiner since it is determined by ${\beta _0}(n)$. But the loss power can be maintain in the accept range by choosing a proper input coupling coefficient. The amount of inputs $n$ may vary from one to dozens, and thus the output coupler must be tunable in a large range. So the waveguide coupler should be taken into consideration \cite{lab10,lab11}.

 On the basis of the above analysis, compared with the first method, the second method is more convenient when the amount of input couplers changes. Therefore, the second method is adopted for designing the high power cavity combiner.

The relation between the input and output coupler coupling coefficients that are calculated with different values of power loss ratio ${\eta _l}$ varying from 1\%-5\% is presented in Fig. \ref{fig3}. From this figure, it can be clearly seen that for a given constant ${\beta _i}$, we can obtain the ranges of combining number and power loss ratio, which are convenient for designing the cavity combiner.

\begin{center}
\includegraphics[width=8cm]{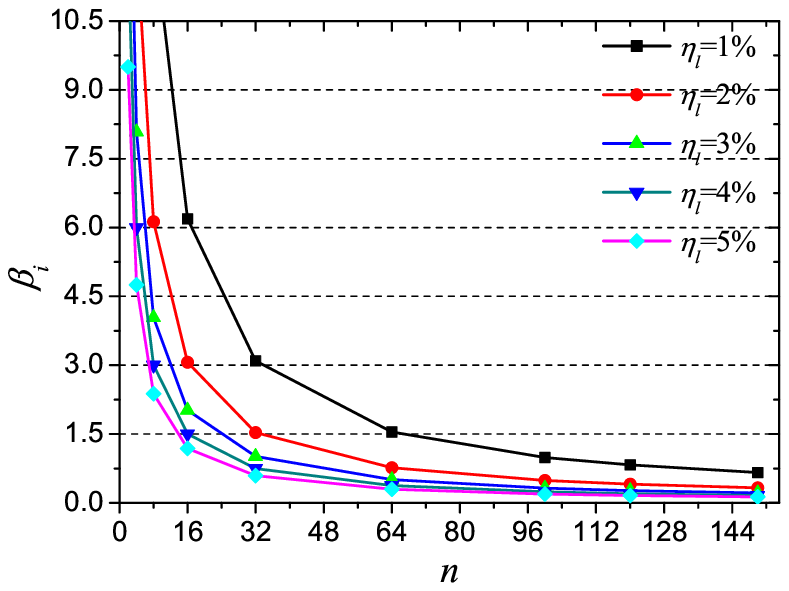}
\figcaption{\label{fig3}  (color online) The change of ${\beta _i}$ with input ways $n$ at different power loss ratios ${\eta _l}$. }
\end{center}


\section{Application}		
To demonstrate the method, we carried out numerical research on a 500 MHz cavity combiner prototype design using CST MWS. The coefficient ${\beta _i}$ is set to 1.5, referring to Fig. \ref{fig3}, which corresponds to the ranges of $n$ and ${\beta _0}$ are 16-64, 20-100, respectively. The specific parameters of the cavity combiner prototype are listed in Table~\ref{tab1}.

\begin{center}
\tabcaption{ \label{tab1}  The parameters of the cavity combiner prototype.}
\footnotesize
\begin{tabular*}{80mm}{c@{\extracolsep{\fill}}ccc}
\toprule
frequency/MHz                   &500\\
max. output power/kW                         &80\\
input power per way/kW       &1.5\\
range of the input ways n            &16, 32, 48, 64\\
input coupling coefficient ${\beta _i}$            &1.5\\
output coupling coefficient ${\beta _0}$           &20-100\\

\bottomrule
\end{tabular*}
\vspace{0mm}
\end{center}
\vspace{0mm}

\subsection{Resonant cavity and couplers design}		
The layout of our designed cavity combiner operating at 500 MHz is shown in Fig. \ref{fig4}. The cavity is made of aluminium and the cross-section of the cavity is regular hexadecagon. The sixteen side walls of the cavity are assembled and disassembled, eight of which will totally employ 64 input couplers. The TE${_{010}}$ mode is chosen as the operating mode and the radius of the incircle of 16 polygon is ${R_0} = {{2.405 * {\lambda _0}} \mathord{\left/
 {\vphantom {{2.405 * {\lambda _0}} {\left( {2 * \pi } \right)}}} \right.
 \kern-\nulldelimiterspace} {\left( {2 * \pi } \right)}}$ \cite{lab12}. To make sure that there is enough bandwidth between TE${_{010}}$ and other modes, the height of the cavity must be set to a proper value.

\begin{center}
\includegraphics[width=8cm]{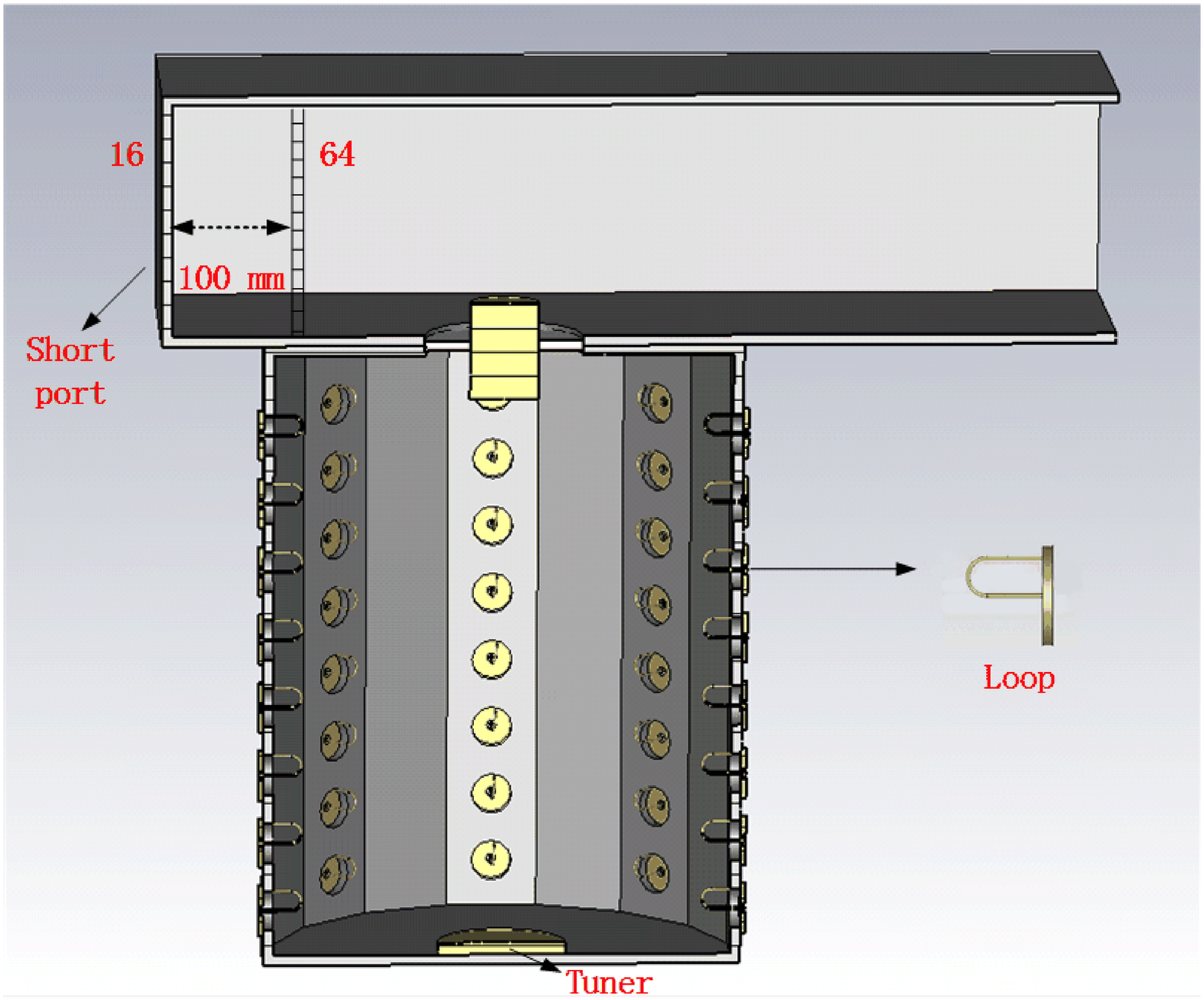}
\figcaption{\label{fig4}  The layout of the cavity combiner. }
\end{center}

Magnetic-coupling loops mounted on the cavity are used to feed energy into the cavity which is also shown in Fig. \ref{fig4}. The coupling coefficient can be adjusted by attaching a rotatable flange on coupler. The dimensions of the loop that can be estimated by the analytical equation in the reference \cite{lab13} and then the coupling coefficient of the input couplers is optimized to 1.5.

A WR 1800 waveguide transition to 6 1/8 inch coaxial used as the output coupler is shown in Fig. \ref{fig4}. The short port of the waveguide is shiftable, which has a large effect on tuning the coupling coefficient ${\beta _0}$. To guarantee that the coupling coefficient can be varied between 15 and 110, which needs some allowance, the length of probe and the shiftable short port of the waveguide are calculated and chosen by CST MWS.

\subsection{Simulation results}		
The numerical study on the cavity combiner is carried out from 16 to 64 input ways, which increases in step of 16.

There are 16 same input couplers and output coupler optimized above that are symmetrically mounted on the cavity by columns, and each side wall employed with 8 couplers. Before simulation, we define the output waveguide as 1 port, the input couplers as 2-17 port by columns. Due to the perturbation caused by couplers, the resonant frequency of the cavity may deviate from the design value, so the tuner is needed, which is inserted into the bottom of the cavity as shown in Fig. \ref{fig4}. The calculated results of the S parameter of 16 ways cavity combiner are given in the Fig. \ref{fig5}. We can see that the scattering coefficient S${_{11}}$ is better than -30 dB at 500 MHz, which indicates the combiner is matched by tuning the short port of the waveguide and the tuner.

\end{multicols}

\begin{center}
\includegraphics[width=14cm]{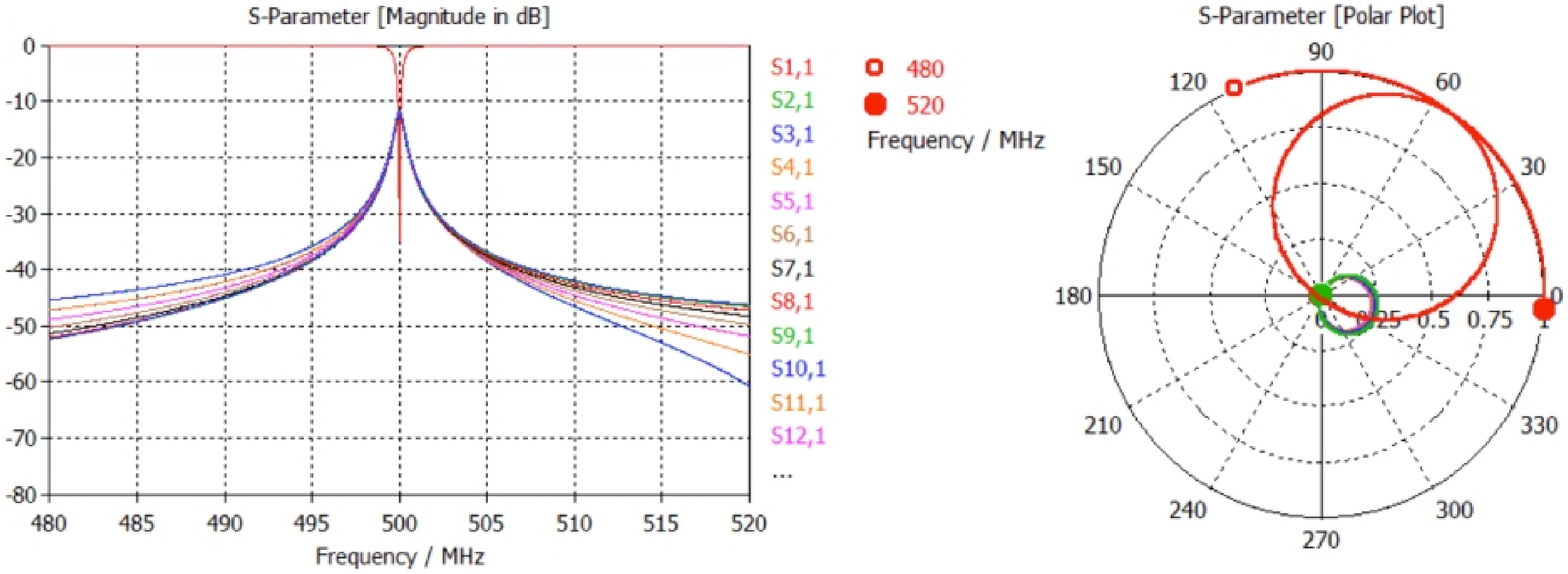}
\figcaption{\label{fig5}  (color online) The S parameters of 16 ways matching cavity combiner. }
\end{center}

\begin{multicols}{2}

Besides, the cases of 32, 48, 64 input couplers symmetrically mounted on the cavity are studied in the same way as in the case of 16 input couplers above. By means of adjusting the short port of the waveguide and the tuner, the S${_{11}}$ of 32, 48 and 64 ways cavity combiner is tuned less than -30 dB at resonant frequency 500 MHz, which means that the combiner achieves at the matching condition. As compared with 16 ways, in the case of 64 ways the short way of the waveguide is shifted 100 mm along the waveguide transmission direction under the matching condition. The matching scattering coefficients of 64 ways cavity combiner are given in Fig. \ref{fig6}. Fig. \ref{fig7} shows the change of the maximum, minimum input reflectance coefficients in the simulation and the theoretical reflectance coefficients, which are calculated by Eq.~(\ref{eq:gama}), with the amount of input ways. A representative of transmit coefficient S${_{15}}$ and its theoretical values calculated by Eq.~(\ref{eq:Si0}) in the simulation varying along the amount of input ways is shown in Fig. \ref{fig8}.

It can be seen from Fig. \ref{fig7} that the difference between the simulated maximum and minimum ${\Gamma _i}$ is less than 0.1 dB. This is because the TE${_{010}}$ field strengths at different coupling loops are slightly different. We can also see that the theoretical ${\Gamma _i}$ always keep in the range between the simulated maximum and minimum ${\Gamma _i}$, which illustrates all the input ways with good RF coherence and the simulation results were consistent with the theoretical results. We can see from the Fig. \ref{fig8} that the simulated transmit coefficient S${_{15}}$ are also consistent with the theoretical values.

\end{multicols}

\begin{center}
\includegraphics[width=14cm]{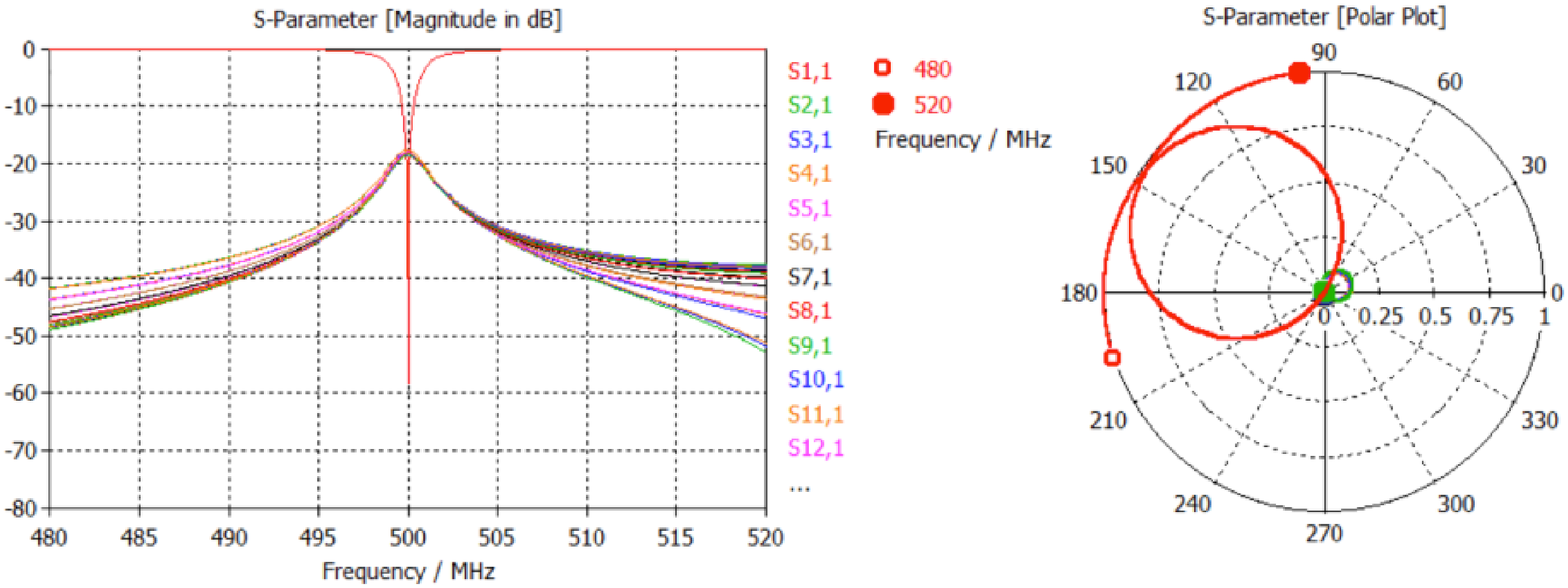}
\figcaption{\label{fig6}  (color online) The S parameters of 64 ways matching cavity combiner. }
\end{center}

\begin{multicols}{2}

All the simulation results above demonstrate that the cavity designed using the proposed method is capable of combining 16-64 same input ways, just needing to tune the short port of the waveguide and tuner when the number of input ways changes, and the results are in good agreement with the proposed method, which means that the design method of cavity combiner is feasible.
\begin{center}
\includegraphics[width=7cm]{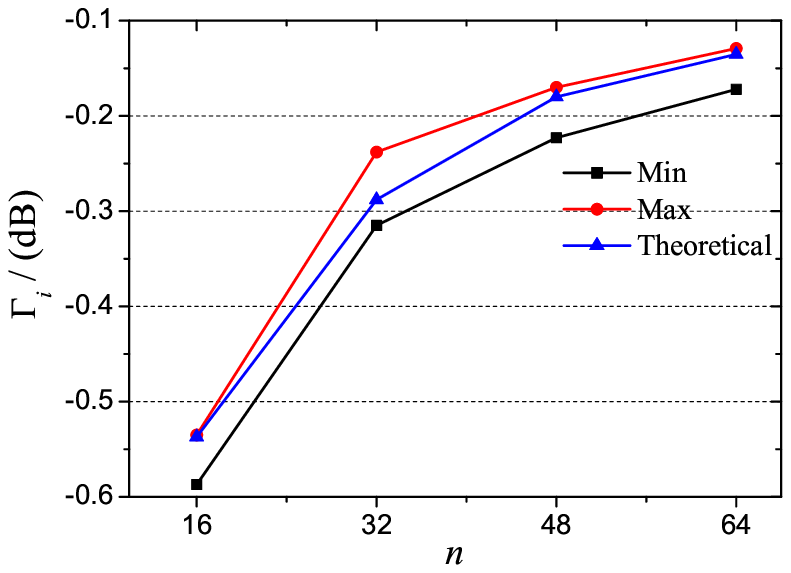}
\figcaption{\label{fig7}  (color online) The change of ${\Gamma _i}$ with the amount of input ways. }
\end{center}

\begin{center}
\includegraphics[width=7cm]{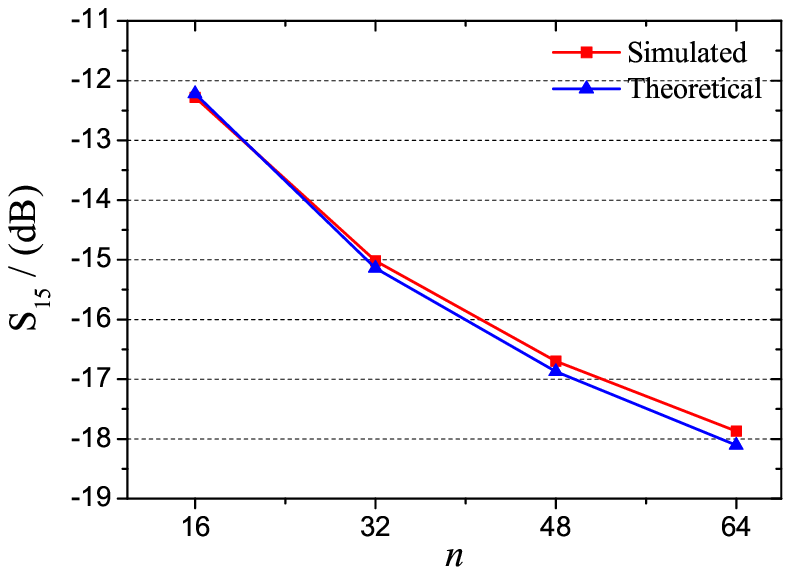}
\figcaption{\label{fig8}  (color online) The change of S${_{15}}$ with the amount of input ways.}
\end{center}

\section{Conclusion}		

In this paper, we first analyzed the equivalent circuit of the cavity combiner, and derived the relation between the input and output coupling coefficients, as well as the scattering parameters under the matching condition. Based on the obtained relation and practical application, it is suggested that the range of power loss rate for the maximum outputting power is 0.5\%-1\% and the number of input ways of the high power cavity combiner is less than 200. Then we proposed a method for designing the variable-way high-power cavity combiner, in which the input coupling coefficient is constant and the output coupling coefficient is tunable. The method was applied to the design of a cavity combiner with 500 MHz and variable input ways from 16 to 64, and the simulation results agree well with the theoretical  RF parameters, thus demonstrating the reliability of the proposed method. In addition, to experimentally test and verify the simulation results, a prototype is being designed and will be manufactured.

\end{multicols}

\vspace{-1mm}
\centerline{\rule{80mm}{0.1pt}}
\vspace{2mm}

\begin{multicols}{2}

\end{multicols}

\clearpage
\end{CJK*}

\begin{thebibliography}{90}

\vspace{3mm}



\bibitem{lab1} P. Marchand, T. Ruan, F. Ribeiro et al., Phys. Rev. ST Accel. Beams, {\bf10}: 112001 (2007)

\bibitem{lab2} P. Marchand, M. Louvet, K. Tavakoli et al., Successful RF and Cryogenic Tests of the SOLEIL Cryomodule, in \emph{Proceedings of the 21st Particle Accelerator Conference}, (Tennessee: PAC'05 ORNL/SNS, 2010) p. 3438

\bibitem{lab3} J. Jacob, J.-M. Mercier, M. Langlois et al., 352.2 MHz-150 KW Solid State Amplifiers at The ESRF, in \emph{Proceedings of the 2nd International Particle Accelerator Conference}, (San Sebastian: IPAC'11 EPS-AG, 2011), p. 71

\bibitem{lab4} Bravo, B, Mares, F and Perez et al., CACO: A CAVITY COMBINER FOR IOT AMPLIFIERS, in \emph{Proceedings of the 2nd International Particle Accelerator Conference}, (San Sebastian: IPAC'11 EPS-AG, 2011) p. 181


\bibitem{lab5} J. Jacob, L. Farvacque, G. Gautier et al., Commissioning of First 352.2 MHz-150 kW Solid State Amplifiers at the ESRF and Status of R\&D, in \emph{Proceedings of the 4th International Particle Accelerator Conference}, (Shanghai: IPAC'13 OC/SPC, 2013) p. 2708

\bibitem{lab6} CST Microwave Studio manual, 2011, http://www.cst.com

\bibitem{lab7}  Status of RF Solid State Amplifier Project, http://www.crisp-fp7.eu/fileadmin/\_migrated/content\_uploads/CRISP\_AM\_2012
\_WP7\_Jacob.pdf, (2012) 1-13

\bibitem{lab8}  Thomas P. Wangler, \emph{RF linear accelerators} (2nd edition, John Wiley \& Sons, Inc., 2008), p. 145

\bibitem{lab9}  David M. Pozar, \emph{Microwave engineering} (Fourth edition, John Wiley \& Sons, Inc., 2009)

\bibitem{lab10}  Rajesh Kumar, P. Singh, Divya Unnikrishnan et al., Nucl. Instrum. Methods A, {\bf481}: 203-213 (2012)

\bibitem{lab11}  J. Gao, Nucl. Instrum. Methods A, {\bf664}: 36-42 (2002)

\bibitem{lab12}  V. Khoruzhiy, Gyromonotron as an instrument for electron beam cooling, in accelererators, arXiv:1305.3770, (2013)

\bibitem{lab13}  J. Corlett, Derun Li, A. Mitra, A 35 MHz re-buncher RF cavity for ISAC at Triumf, in \emph{Proceedings of the 6th  European Particle Accelerator Conference}, (Stockholm: EPAC'98 IOP/Ltd, 1998), p. 1790


\end{thebibliography}
\end{document}